\documentclass[usenatbib,useAMS]{mn2e}

\newcommand{\apj}{ApJ}
\newcommand{\aap}{A\&A}
\newcommand{\apjl}{ApJL}
\newcommand{\mnras}{MNRAS}
\newcommand{\nat}{Nature}

\usepackage{color}
\usepackage{graphicx}
\usepackage{amsmath}
\usepackage{amssymb}
\usepackage{perpage} 
\MakePerPage{footnote} 

\title[OH depletion and the mass-to-flux ratio]{Effect of OH depletion
  on measurements of the \\ mass-to-flux ratio in molecular
  cloud cores}
\author[K. Tassis et al.]
{Konstantinos Tassis$^{1,2}$\thanks{tassis@physics.uoc.gr}, Karen Willacy$^{3}$, Harold
  W. Yorke$^{3}$, Neal J. Turner$^{3}$\\ 
$^{1}$Department of Physics and ITCP\thanks{Institute of Theoretical
  and Computational Physics (formerly Institute of Plasma Physics)}, University of Crete, 71003, Heraklion, Greece\\
$^{2}$Foundation for Research and Technology - Hellas, IESL, Voutes, 7110 Heraklion, Greece\\
$^{3}$Jet Propulsion Laboratory, California Institute of Technology, Pasadena, CA 91109, USA
}
\begin{document}



\maketitle

\label{firstpage}

\begin{abstract}
The ratio of mass and magnetic flux  determines the relative importance of magnetic and gravitational forces in the evolution of molecular clouds and their cores. Its measurement is thus central in discriminating between different theories of core formation and evolution. Here we discuss the effect of chemical depletion on measurements of the mass-to-flux ratio using the same molecule (OH) both for Zeeman measurements of the magnetic field and the determination of the mass
of the region. The uncertainties entering through the OH
abundance in determining separately the magnetic
field and the mass of a region have been recognized in the
literature. It has been proposed however that, when comparing two regions of the same cloud, the abundance will in both cases be the same. We show that this
assumption is invalid. We demonstrate that when comparing
regions with different densities, the effect of OH depletion in measuring changes of the mass-to-flux ratio between different parts of the same cloud can even reverse
the direction of the underlying trends (for example, the mass-to-flux
ratio may appear to {\em decrease} as we move to higher density regions). The systematic errors enter primarily through the inadequate estimation of the mass of the region. 

 \end{abstract}

\begin{keywords}
ISM: clouds -- ISM: magnetic fields -- ISM: abundances -- ISM: molecules -- MHD -- stars: formation
\end{keywords}

\section{Introduction}\label{intro}

Identifying the mechanism that regulates the star formation process and keeps the
star formation efficiency low is at the heart of current theoretical
and observational efforts. Magnetic fields and turbulence are the two main
contestants for the role of the agent that (a) supports molecular
clouds against gravity so that they are not globally collapsing and
(b) mediates cloud fragmentation into cores that will later go on to
form protostars. Despite several
decades of theoretical study and observations, which of the two
mechanisms is dominant in molecular clouds is still a matter of heated
debate. The reason is that the two appear to be in rough
equipartition in cold molecular clouds (Heiles \& Troland 2005), and observational degeneracies and
uncertainties have thus far thwarted efforts to declare a clear
winner.

The ratio between mass and magnetic flux (the {\em mass-to-flux
  ratio}, $M/\Phi_B$) is a quantity of central importance in moving this debate
forward, for two reasons. First, it quantifies the effectiveness of the magnetic field
alone in supporting a cloud against gravity: if the mass-to-flux ratio
exceeds a critical value
\begin{equation}\label{crit}
\left(\frac{M}{\Phi_B}\right)_{\rm crit} = \left(\frac{1}{63G}\right)^{1/2}\,,
\end{equation}
(Mouschovias \& Spitzer 1976), then the magnetic field cannot support
the region in question against its self-gravity and, in absence of any other
means of support, the region will collapse dynamically. Such a region
is called {\em magnetically supercritical}. If on the other hand the
mass-to-flux ratio is smaller than the critical value, then magnetic
forces alone can support the region against its self-gravity. Such a
region is called {\em magnetically subcritical}. 

If molecular clouds are magnetically subcritcal as a whole, then in
order for gravitational collapse to take place, magnetically
supercritical fragments have to first form within them. This is
achieved through the process of ambipolar diffusion: because of the
imperfect coupling between magnetic fields and the neutral gas in
weakly ionized molecular clouds, neutral molecules can move
diffusively through ions and magnetic field lines and increase the
mass-to-flux ratio around centers of gravity. In this way,
supercritical cores can form within subcritical clouds (e.g., Fiedler
\& Mouschovias 1993.) In order for a molecular cloud core to be
observable as a distinct fragment, it typically is already
supercritical (e.g. Tassis \& Mouschovias 2004.)

This mechanism of core formation leaves a characteristic imprint on
the relative mass-to-flux ratio between supercritical core and
subcritical envelope: the core always has a higher mass-to-flux ratio
than its parent cloud envelope. Crutcher et al. (2009) took advantage
of this property to design a test of ambipolar diffusion as a core
formation mechanism. They attempted to measure, through OH Zeeman
splitting, the magnetic field of cloud envelopes in locations around
four cores with previously measured magnetic fields. They additionally
used the integrated intensity of OH to estimate the mass in cores and
envelopes. Their quantity of interest was the ratio $R'$ between the
mass-to-flux ratio in the core over the mass-to-flux ratio in a larger
area, enclosing both core and envelope, which ambipolar-diffusion--driven
fragmentation predicts should be higher than or equal to one. 

Crutcher et al. (2009) argued that, because the quantity they are
trying to determine is a ratio between the same quantities in
different regions of a single cloud, unknown parameters, such as the
orientation of the magnetic field with respect to the line of sight
and the OH abundance, drop out. The uniform orientation assumption, as
well as the particulars of the Crutcher et al. (2009) implementation
(selection of locations of Zeeman observations in the cloud envelopes,
statistical treatment of upper limits) have been criticized in the
literature (Mouschovias \& Tassis 2009, 2010.) Chemistry can also affect such an experiment because
the OH abundance and the magnetic field are convolved in a Zeeman measurement
(Tassis et al. 2012b). However, we show here 
that the biggest source of systematic errors in such experiments is the effect of chemistry on the mass estimate.  

Briefly stated, the chemical process
of OH depletion onto grains invalidates the assumption of a common OH
abundance throughout the cloud. Depletion
makes the OH abundance lower in higher
density regions, leading to an underestimate of their total mass and
consequently an underestimate of their mass-to-flux ratio. As a
result, the mass-to-flux ratio of a supercritical fragment can appear
{\em lower} than that of the subcritical envelope, {\em even if all other
assumptions hold, and all other parameters and analyses of the
experiment are exactly correct. }

Here we quantify this effect, using chemodynamical MHD models
that follow concurrently the evolution of density, magnetic field, and
chemical abundace in a forming and evolving molecular cloud core. We
trace the origin of the effect in the radial
dependence of mass and magnetic field in 
ambipolar-diffussion--produced cores within molecular clouds of
various initial parameters; and we suggest a straightforward way to
overcome this systematic error: use of {\em dust continuum emission}
rather than OH intensity to measure the masses that enter into
mass-to-flux ratio estimates. 

This paper is organized as follows. In \S \ref{models} we briefly
review the suite of chemodynamical models used in this work. We use
these models in \S \ref{obs} to construct simulated observations which
demonstrate the effect of chemistry on estimates of mass and the
mass-to-flux ratio in simulated cores. Finally, we summarize and discuss our conclusions in \S
\ref{discussion}. 

\section{Chemodynamical Models}\label{models}

For the simulated observations presented here, we use the suite of
1.5D magnetic chemodynamical models described in detail in Tassis et
al. (2012a) and used in Tassis et al (2012b) to study the effect of OH
depletion on the measurement of the magnetic field. We briefly review
their properties below. 

We use results from 14 models of evolving prestellar cores that couple
nonequlibrium chemistry with nonideal magnetohydrodynamics. The parameters that are varied are the
initial value of the mass-to-flux ratio; the initial elemental C/O
ratio; the cosmic-ray ionization rate; and the cloud temperature. 

Our ``reference'' model has a  mass-to-flux ratio equal to the critical value for collapse, a temperature of 10 K, a C/O ratio of 0.4, and a cosmic ray ionization rate of   $\zeta = 1.3 \times 10^{-17} {\rm \, s^{-1}}$.  We examine two additional values of the initial mass-to-flux ratio: 1.3 times the critical value 
(a faster-evolving, magnetically supercritical model), and 0.7 of the critical value (a slower, 
magnetically subcritical model). For each of these three dynamical
models, the carbon-to-oxygen ratio is varied from its ``reference''
value by keeping the abundance of C constant and changing that of
O. The other value of C/O ratio examined is 1. 
In addition, to test the effect of the temperature, we have varied each of the three basic dynamical models by changing $T$ by a factor of  $\sim 1.5$ from its reference value of 10 K and examined models with $T = 7$ K and $T = 15$ K. 
Finally, to test the effect of the cosmic ray ionization rate, we have studied two additional models, which have a ``reference'' value for the temperature, C/O ratio, and mass-to-flux ratio, but for which $\zeta$ is varied by a factor of four above ($\zeta = 5.2 \times 10^{-17}$ $s^{-1}$) and below ($\zeta = 3.3 \times 10^{-18}$ $s^{-1}$) its ``reference'' value (covering the range of observational estimates \cite[e.g.][]{McC2003,Hez2008}.

\section{Simulated Observations}\label{obs}

\begin{figure}
\includegraphics[width=8cm, clip=true]{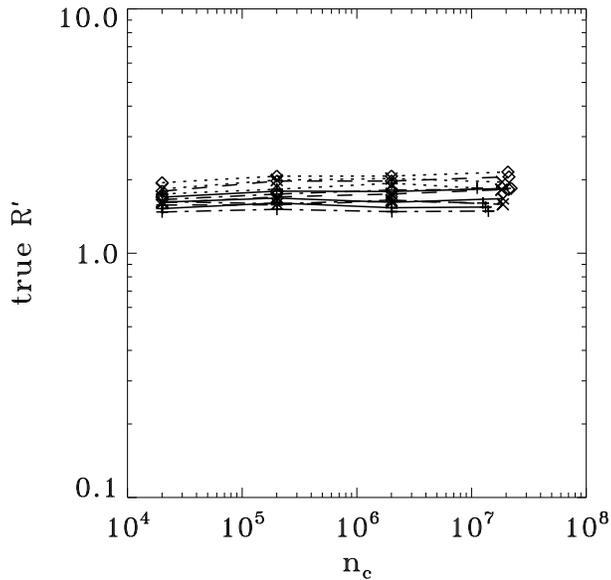}
\caption{\label{real}
Actual value of $R'=(M/\Phi_4)/M/\Phi_3$ as a function of central density of
the core, for the various models used in this
work. Models correspond to symbols and line types as follows: Diamonds: magnetically subcritical; X: magnetcially
critical; crosses: magnetically supercritical; solid lines: reference
C/O, T, $\zeta$; dotted lines: C/O=1; dashed lines: low T; dot-dashed:
high T; thick line: high $\zeta$; thin line: low $\zeta$.}
\end{figure}
 
We first use the models discussed in \S \ref{models} to construct an
idealized experiment similar to the one proposed by Crutcher et
al. (2009). Because of the presence of an ordered magnetic field, each
forming core preferentially collapses along field lines (where there
is no magnetic support), thus assuming a disk shape. We assume
that the disk is observed face-on. We define a ``dense core'' in each model to be a disk extending out to
the radius where the number density of $H_2$ is equal to 10$^4{\rm \,
  cm^{-3}}$, and a ``core+envelope'' region to be a disk extending out
to where $n_{\rm H_2} = 10^{3} {\rm \, cm^{-3}}$. For each model, and
at various stages of evolution, we measure the mass-to-flux ratios of
``dense core'', $M/\Phi_4$, and ``core+envelope'', $M/\Phi_3$, using the exact quantities for
mass and magnetic flux recorded in our simulations. We then plot their
ratio, $R'=(M/\Phi_4)/(M/\Phi_3)$ as a function of evolutionary
stage. The latter is quantified by the value of the central density in
the core at every time instant. These results are shown in
Fig. \ref{real}. The ratio between mass-to-flux ratios is slightly above unity for all
models, but no higher than 2. As expected (e.g. Fiedler \& Mouschovias 1993 {\bf
  ?}), there is essentially no evolution of the
mass-to-flux ratio of the cores during these stages of collapse: 
the envelope is magnetically supported and neither its mass
nor its magnetic flux evolve in time; the mass-to-flux ratio of the
core is also steady after 
the onset of dynamical collapse, and before the ``resurrection'' of
ambipolar diffusion, seen at densities around $10^{10} {\rm \,
  cm^{-3}}$ (Desch \& Mouschovias 2001; Tassis \& Mouschovias 2007.)

\begin{figure}
\includegraphics[width=8cm, clip=true]{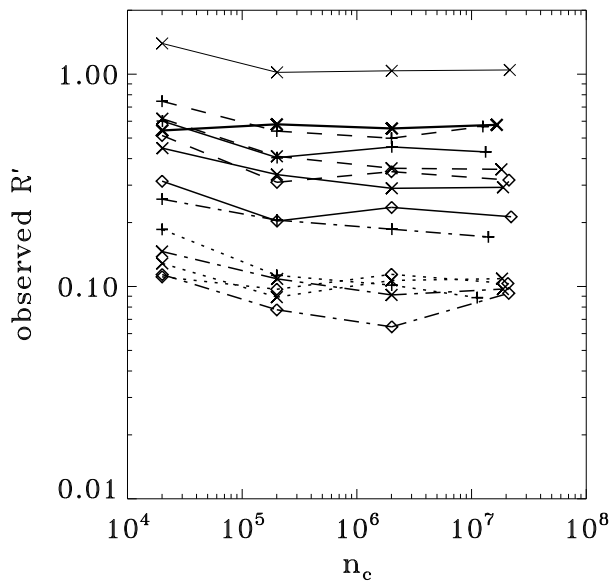}
\caption{\label{fobs}
Simulated observations of  $R'=(M/\Phi_4)/M/\Phi_3$ as a function of central density of
the core, for the various models used in this work. Symbols and lines
as in Fig.~\ref{real}.}
\end{figure}

Next, we use our simulations to reproduce what a ``realistic'' 
experiment would look like: we calculate the mass and magnetic flux that would be
estimated for the ``dense core'' and ``core+envelope'' regions using Zeeman
observations and observations of the OH line intenstity
respectively. To do this, we assume that the magnetic field traced by
Zeeman splitting of the OH line is the OH-mass--weighted magnetic
field in the region under consideration. Since our models follow both
the OH abundance and the evolution of the magnetic field, this quantity
is easily calculated. In addition, we assume that
the cloud is optically thin in OH line emission, and therefore the OH
intensity is proportional to the OH column density. The ratio of
masses between core and core+envelope (which enters the estimation of
$R'$) is thus simply the ratio between OH masses between the two
regions, a quantity that is directly followed in each model. 

OH depletion affects the measurement of both the magnetic field and
the mass {\em in the same direction}, so it is not {\it a priori}
obvious whether $R'$ would end up underestimated or overestimated. The
mean magnetic field is underestimated, as discussed in Tassis et
al. (2012), because OH is more depleted in the higher density regions,
where the magnetic field is also high. The total mass in the highest
density regions also appears lower in OH intensity, because the OH
abundance drops as the density increases. 

Fig.~\ref{fobs} shows the ratio $R'$ as measured in the fashion
discussed above, plotted against the central number
density (a proxy for the evolutionary age), for different models. In all models $R'$ is underestimated
compared to its ``real'' value. In all models but one, its
``observed'' value is lower than one, and, depending on the model, it
can appear to be lower than $0.1$. Ironically, the more subcritical an
envelope is, the more $R'$ deviates from its ``real'' value: the
reason is that more subcritical models evolve more slowly and the
effect of depletion is more severe.

\begin{figure}
\includegraphics[width=8cm, clip=true]{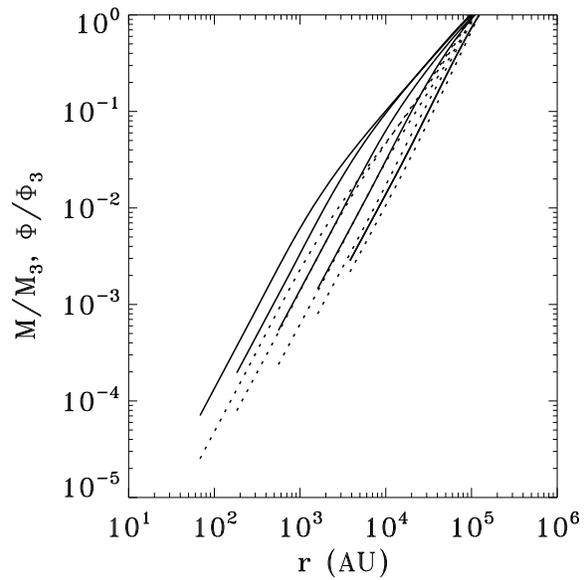}
\caption{\label{profiles}
Scaling of the total (cumulative) mass (solid lines) and magnetic
flux (dotted lines) with radius, for a region of the cloud with an
outer boundary at $10^3 {\rm \, cm^{-3}}$. Each curve corresponds to a different
evolutionary stage.
}
\end{figure}

The reason why the net effect of depletion has the sign discussed
above can be traced back to the cumulative radial profiles of the total mass and
total flux in the evolving cloud. We plot these profiles in
Fig.~\ref{profiles}. The total mass (solid lines) and magnetic flux
(dotted lines) are normalized to unity
at a radius where the number density is $10^3 {\rm \, cm^{-3}}$.
As expected for a disk-like structure,
the scaling of mass and flux with radius have the same slope in the
inner parts of the disk. However, the overall normalization of the
scaling evolves differently in time. Because of OH depletion, it is
the outermost regions that dominate the ``observed'' values of both
mass and flux. Because of the different relative distributions of
total mass and flux with radius, the ``observed'' mass is more
severely affected than the ``observed'' flux, and the mass-to-flux
ratio appears underestimated, more so for higher-density regions. As a
result, measured values of the ratio $R'$ tend to be lower than the true ones
obtained using OH to estimate the core and envelope mass. 

\section{Summary and Discussion}\label{discussion}

We have used a suite of chemodynamical simulations to demonstrate the
effect of OH depletion on observational estimates of the
mass-to-flux ratios in molecular cloud cores and envelopes,
when OH intensity is used as means to estimate the mass associated
with the observed region. 

We have shown that, whereas OH depletion leads to the underestimation of
both mass and flux in the densest regions, the effect on the mass is
more severe, leading to an overall {\em underestimate} of the
mass-to-flux ratio. The effect is more significant in the denser
regions, where depletion is more severe. As a result, the mass-to-flux
ratio can appear to {\em decrease} as we move from a subcritical cloud's envelope
to its supercritical ambipolar-diffusion--formed core - despite the
fact that the mass-to-flux ratio is in reality always
higher in the core. 

Our result is a sobering reminder that chemical effects can be the
source of severe systematic errors, for all
experiments setting out to measure the mass-to-flux ratio in molecular
cloud cores and envelopes. The most severe effect is that of tracer
depletion on the measured mass. The effect of depletion has been
observationally confirmed in several prestellar cores
(e.g., Caselli et al. 1999; Tafalla et al. 2002; Tafalla et
al. 2004), including L1544 (one of the cores studied by Crutcher et
al. 2009). 


The effect of depletion in estimates of the magnetic field, on the other hand, needs to be
carefully accounted for (see also Tassis et al.~2012). As a general
rule-of-thumb, Zeeman observations of a region centered on a center of
gravity will always preferentially probe the outermost (least dense)
parts of the region, where OH (or other tracer) depletion is least
severe. 

Regarding experiments specifically designed to study the gradient of
the mass-to-flux ratio between cores and envelopes (e.g. Crutcher et
al. 
2009), we point out that the effect of depletion alone is
  enough to make the mass-to-flux ratio in the core appear smaller
  than that of the envelope. Because depletion is more severe in
  clouds that evolve more slowly, {\em the decrease of the mass to flux
  ratio from envelope to core will be greater when the
  magnetic field is more dynamically important.}

Crutcher et al. (2009) have claimed that they have constrained $R'$ to
be lower than 1 in four cases of molecular cloud cores and the
envelopes surrounding them. Although their statistical analysis and
selection of envelope locations to observe have been challenged
(Mouschovias \& Tassis 2009, 2010), our findings have important
implications for these results, {\em even if taken at
  face-value}. Crutcher et al. (2009) used the assumption of a
constant OH abundance throughout core and envelope to estimate the
ratio of masses that enters into their estimation of $R'$, just as we have
done in our simulated observations. Therefore, any  
values of $R'$ that they find to be much smaller than 1 (even if one was to otherwise accept
their analysis) do not imply that a core was not
formed by ambipolar diffusion. 

In conclusion, the assumption of a constant OH
abundance, even within a small region of the same molecular cloud,
is a poor assumption when there is a density gradient in the
region under consideration. Nevertheless, experiments measuring the
gradient of mass-to-flux ratio between core and envelope can
be important tests
of the predictions of ambipolar diffusion theory, and for this reason
it is imperative that they are performed carefully. In the context of
this work, such experiments require the accurate determination of the
core mass independent from the OH
lines used for Zeeman observations, preferably using dust continuum
observations. We note however that the dust-to-gas ratio can itself vary with density:
Ciolek \& Mouschovias (1994) found that, because of the effect on ambipolar
diffusion on ionized grains, the abundance of grains in the
supercritical core is reduced relative to that in the envelope by a
factor that is equal to that by which the envelope is initially
subcritical. This effect should therefore be corrected for in such
measurements. 


\section*{Acknowledgments}

We thank Telemachos Mouschovias and Vasiliki Pavlidou for helpful
discussions. K.T. is acknowledging support by FP7 through Marie
Curie Career Integration Grant PCIG-GA-2011-293531 ``SFOnset'' and the EU FP7 Grant
PIRSES-GA-2012-31578 ``EuroCal''.  This work was partially supported by the ``RoboPol'' project, which is
implemented under the ``ARISTEIA'' Action of the ``OPERATIONAL PROGRAMME EDUCATION AND
LIFELONG LEARNING'' and is co-funded by the European Social Fund (ESF) and Greek National
Resources. The project
was supported in part by the NASA Origins of Solar Systems
program. Portions of this research were conducted at the Jet 
Propulsion Laboratory, California Institute of Technology, which is 
supported by the National Aeronautics and Space Administration
(NASA).  \copyright 2014. All rights reserved.


\begin{thebibliography}{30}

\bibitem[Caselli et al.(1999)]{1999ApJ...523L.165C} Caselli, P., Walmsley, 
C.~M., Tafalla, M., Dore, L., \& Myers, P.~C.\ 1999, \apjl, 523, L165 

\bibitem[Ciolek 
\& Mouschovias(1994)]{1994ApJ...425..142C} Ciolek, G.~E., \& Mouschovias, T.~Ch.\ 1994, \apj, 425, 142 


\bibitem[Crutcher et al.(2009)]{cht} Crutcher, R.~M., 
Hakobian, N., \& Troland, T.~H.\ 2009, \apj, 692, 844 

\bibitem[Desch 
\& Mouschovias(2001)]{2001ApJ...550..314D} Desch, S.~J., \& Mouschovias, T.~Ch.\ 2001, \apj, 550, 314 


\bibitem[Evans et al.(2001)]{2001ApJ...557..193E} Evans, N.~J., II, 
Rawlings, J.~M.~C., Shirley, Y.~L., \& Mundy, L.~G.\ 2001, \apj, 557, 193 

\bibitem[Fiedler 
\& Mouschovias(1993)]{1993ApJ...415..680F} Fiedler, R.~A., \& Mouschovias, T.~Ch.\ 1993, \apj, 415, 680 

\bibitem[Heiles 
\& Troland(2005)]{2005ApJ...624..773H} Heiles, C., \& Troland, T.~H.\ 2005, \apj, 624, 773 

\bibitem[Hezareh et al.(2008)]{Hez2008}
 Hezareh, T., Houde, M., 
McCoey, C., Vastel, C., \& Peng, R.\ 2008, \apj, 684, 1221 

\bibitem[McCall et al.(2003)]{McC2003} 
McCall, B.~J., Huneycutt, A.~J., Saykally, R.~J., et al.\ 2003, \nat,
422, 500 

\bibitem[Mouschovias 
\& Spitzer(1976)]{1976ApJ...210..326M} Mouschovias, T.~Ch., \& Spitzer,
L., Jr.\ 1976, \apj, 210, 326 

\bibitem[Mouschovias 
\& Tassis(2009)]{mt09} Mouschovias, T.~Ch., \& Tassis,
K.\ 2009, \mnras, 400, L15 

\bibitem[Mouschovias 
\& Tassis(2010)]{mt10M} Mouschovias, T.~Ch., \& Tassis, K.\ 2010, \mnras, 409, 801 

\bibitem[Tafalla et al.(2002)]{2002ApJ...569..815T} Tafalla, M., Myers, 
P.~C., Caselli, P., Walmsley, C.~M., \& Comito, C.\ 2002, \apj, 569, 815 

\bibitem[Tafalla et 
al.(2004)]{2004A&A...416..191T} Tafalla, M., Myers, P.~C., Caselli, P., \& Walmsley, C.~M.\ 2004, \aap, 416, 191 

\bibitem[Tassis 
\& Mouschovias(2004)]{2004ApJ...616..283T} Tassis, K., \& Mouschovias,
T.~Ch.\ 2004, \apj, 616, 283

\bibitem[Tassis 
\& Mouschovias(2007)]{2007ApJ...660..388T} Tassis, K., \& Mouschovias,
T.~Ch.\ 2007, \apj, 660, 388

\bibitem[Tassis et al.(2012)]{paper1} Tassis, K., Willacy, K., 
Yorke, H.~W., \& Turner, N.~J.\ 2012a, \apj, 753, 29 

\bibitem[Tassis et al.(2012)]{paper2} Tassis, K., Willacy, K., 
Yorke, H.~W., \& Turner, N.~J.\ 2012b, \apj, 754, 6





\end{thebibliography}
\end{document}